# Synthesis and characterization of ferromagnetic cobalt nanospheres, nanodiscs and nanocubes

D. Srikala[1], V. N. Singh[2], A. Banerjee[3], B. R. Mehta[2] and S. Patnaik[1]

[1] School of Physical Sciences, Jawaharlal Nehru University, New Delhi 110067, India

[2] Thin Film Laboratory, Department of Physics, Indian Institute of Technology Delhi, Hauz Khas, New Delhi 110016, India

[3] UGC-DAE Consortium for Scientific Research, University Campus, Khandwa Road, Indore 452017, India

Corresponding author mail id: spatnaik@mail.jnu.ac.in

## Abstract

We report the synthesis of cobalt nanoparticles with different shapes and sizes by rapid pyrolysis of cobalt carbonyl in the presence of various surfactants. The size and shape of the nanoparticles were influenced by reaction conditions, such as type of the surfactant, molar ratio of surfactant to precursor, reflux temperature and reaction time. The shapes that we have achieved include spherical, nearly spherical, disc and cube. The presence of linear amine yielded nanodiscs and they spontaneously self-assembled into long ribbons. The effect of shape anisotropy on magnetic nanoparticles has been investigated. Spherical nanoparticles of diameter 14.5 nm show strong ferromagnetic behavior at low temperature and superparamagnetism at room temperature. On the other hand the cubic nanoparticles of 45 nm sides showed negligible coercive field at $T = 10$ K and ferromagnetism that persisted above $T = 300$ K. The cobalt nanospheres were oxidized to grow cobalt oxide shell of varying thickness to study exchange bias effect. A pronounced exchange bias and a strong temperature dependant magnetization were observed in oxidized cobalt nanospheres.

**Introduction**

Nanostructured materials have unique electrical, chemical, structural, and magnetic properties because of large volume fraction of surface atoms and because of finite number of atoms in each crystalline core. Recently magnetic nanoparticles have attracted considerable interest due to their promising applications in high-density magnetic recording media[1] and biomedicine[2-3]. All these applications require sustenance of long range order inside each individual nanoparticle at room temperature. Thus the thermal and chemical stability of nanostructured materials at T = 300 K is a critical issue that needs to be addressed for potential technological applications. It is well known that stability can be improved by introducing additional magnetocrystalline anisotropy into the matrix.[4] However, the anisotropy in magnetic nanostructured materials depend not only on the band structure of the parent material, but also on the shape of the nanoparticles and more importantly on exchange coupling across ferromagnetic – antiferromagnetic interfaces.[5-8] Further, the shape of the hysteresis loop is of great importance for magnetic recording applications that require a large remanent magnetization, small coercivity, and (ideally) a square hysteresis loop.

One of the most attractive features of nanostructured magnetic materials is that their magnetic properties can be engineered by the choice of the shape of the constituent nanoparticles.[9-10] Major challenges in this field arise from the fact that most routes used for synthesizing nanoparticles result in assemblies with a wide distribution of particle size and shapes. The method of thermal decomposition of metal carbonyl in hot surfactant solution produces nanoparticles with controllable shape, and narrow size distribution.[11-13]

A wide range of shapes were made by relatively simple variation of monomer concentration.[14-16] The present work explores the use of surfactants in order to yield different shapes and sizes of ferromagnetic cobalt nanoparticles and to study the effect of shape and exchange anisotropy on their magnetic properties. We observe that the shape and size of the cobalt nanoparticles depend sensitively on the type of surfactants used and the reaction time. By optimizing various parameters we have successfully prepared particles of shapes that include spheres, cubes and discs. Further, by exposing to oxygen we could grow antiferromagnetic capping layers on top of bare cobalt spheres. Our results show that the geometry and type of antiferromagnetic capping layer affects more than the surface and volume of contact in controlling the exchange bias.

**Experiments**

Synthesis of Co nanoparticles was carried out using a pyrolysis procedure. Di-cobalt octa-carbonyl $Co_2(CO)_8$ was used as a source for Co atoms. The decomposition of $Co_2(CO)_8$ under inert atmosphere is complex and proceeds through intermediates such as $Co_4(CO)_{12}$, $Co_6(CO)_{16}$ and other unstable mononuclear Co carbonyls. The faster decomposition of $Co_2(CO)_8$ takes place at higher temperatures. Here we employed a combination of surfactants in controlling the shape and size of the nanoparticles. For the synthesis of various shaped and sized Co nanostructures we used the following chemicals. They are Di-cobalt octa-carbonyl $Co_2(CO)_8$ containing 1-5% hexane as a stabilizer, o-dichlorobenzene (DCB, 99 %), oleic acid (OA, 99 %), trioctylphosphine oxide (TOPO, 99%), octadecylamine (ODA, 97%), O,O′-bis(2-aminopropyl)polypropyleneglycol (Jeffamine D-400), and octanoic acid (99.5 %).

The first step for the synthesis of Co nanospheres involved degassing 0.2 g of TOPO with high purity Ar in a three-neck flask for 20 min. Then 0.1 mL of OA + 12 mL of DCB was introduced under Ar atmosphere and heated to reflux temperature (~182 $^o$C). A solution of 0.4 g of $Co_2(CO)_8$ dissolved in 3.6 mL of DCB (precursor solution) was then rapidly injected into the refluxing bath. The solution was allowed to reflux for 15 min and then cooled to room temperature. For the synthesis of nanocubes the DCB was first dried with anhydrous $CaCl_2$ powder and stored under vacuum. A solution containing the surfactants, 0.54 mL O,O′-bis(2- aminopropyl) polypropyleneglycol, 0.09 mL octanoic acid, and 0.12 g TOPO in 13 mL DCB, was degassed for 1 h and then heated to reflux temperature (~186 $^0$C). Then, 0.62 g of $Co_2(CO)_8$ were resolved in 4 mL DCB and rapidly injected in to the reflux bath. The reaction was carried out for 3 min and cooled down with liquid nitrogen. The preparation methodology is summarized in Table 1.

* Table 1

After the reaction, to extract the nanoparticles from the colloidal solution, an equal volume of ethanol and hexane was added and centrifuged for 20 min. Size and shape of Co nanoparticles were studied by high resolution transmission electron microscopy using a technaiG$^2$ (200 KV) microscope; elemental composition study was performed using energy dispersive absorption x-ray spectroscopy (EDAX) and crystal structure was identified using electron diffraction spectroscopy (EDS). Magnetization measurements were carried out in a Quantum Design physical property measurement system in the temperature range from 2 to 350 K and in applied magnetic fields up to 7 Tesla.

**Results and discussion**

Combination of surfactants TOPO and OA yielded spherical shaped *hcp* nanoparticles as shown in fig.1. Fig.1a shows more oxidized nanospheres, 1b shows the electron diffraction pattern, and 1c shows less oxidized Co nanospheres. Our subsequent studies showed that more oxidized nanospheres have $Co_3O_4$ shell and the less oxidized nanospheres have CoO antiferromagnetic shell. With low concentration of TOPO and smaller reaction time an irregular spherical shaped particles were obtained as shown in fig.2a. By replacing one of these surfactants with a linear amine ODA yielded Co nanodiscs. However the distribution in shape was different in the two cases. When TOPO is replaced, ~15 % disc and 85 % spheres are made, where as when OA is replaced, ~ 15 % discs and ~ 85 % cubes are prepared. In prior case the spherical particles are about 11.5 nm and discs are 3.8 nm × 13.5 nm. Oleic acid is well known as a surfactant that adsorbs easily and tightly on the surface of metallic particles with its carboxylic group and greatly impedes the particles to grow in size. This gave rise to monodispersed spherical and disc nanoparticles as shown in fig.2b. But the combination of ODA and TOPO yielded cubic and disc shape nanoparticles as shown in fig.3c. Here TOPO acted as a selective absorber, altering the relative growth rates of different faces of the crystals, as well as a molecule that promotes atom exchange with the monomer during growth of the nanoparticles. The average size of discs formed were 4 nm × 19.3 nm. The role of $-NH_2$ group of amine surfactant was responsible for formation of disc shaped cobalt nanoparticles. Disc yield can be further increased when larger amounts of surfactant are used or when amine is co-injected with the cobalt precursor.[17] Most of the cobalt nanodiscs are self assembled into a ribbon in a way that the larger discs are at the center of the ribbon. Similarly, cobalt nanocubes of varying sizes were prepared with different reaction time by replacing ODA with octanoic acid and O,O′-bis(2-aminopropyl)polypropyleneglycol as shown in fig.3a and 3b. Ideally, a large number of critical nuclei should be formed in a short interval of time when the precursor solution is injected into the hot surfactant mixture. Then a simultaneous and steady growth of those nuclei is expected. Accordingly, 1 min reaction time yielded 30 nm and 3 min reaction time yielded ~ 45 nm sized nanocubes as shown in fig.3a and fig.3b. Single crystalline nature of nanocubes is attested by the diffraction pattern shown fig.3d.

To study the exchange bias effect, the nanospheres were oxidized by exposure to atmospheric oxygen after sonication (to remove the surfactant layer). Exchange bias or unidirectional magnetocrystalline anisotropy provides an extra source of anisotropy energy for particles in holding their magnetic moment against random flipping of spins due to finite size effect of the nanoparticles. When a system with ferromagnetic (FM) and antiferromagnetic (AFM) interface is cooled in the magnetic field through the Néel temperature of AFM lattice, it exhibits unidirectional anisotropy due to the torque exerted by the localized uncompensated AFM spins on the coupled FM spins. So when the particles are cooled in field, a shift in the M-H loop along the field axis generally in the opposite direction to the cooling field is observed, i.e., the absolute value of coercive field for increasing and decreasing field is different. Exchange bias field is measured as $H_E = H_C (FC) - H_C (ZFC)$. A symmetric loop is observed when the sample is cooled in zero field. In our study the more oxidized sample has a typical ferromagnetic Co core of diameter 8.7 nm and antiferromagnetic $Co_3O_4$ shell thickness 2.5 nm and less oxidized sample has 10.6 nm core and 1.9 nm CoO shell thickness. This is shown in insets of

fig.1a and fig.1c respectively. EDAX measurements confirmed the difference in composition ratio of oxygen to cobalt in both the samples. Electron diffraction pattern confirmed the hexagonal close pack structure of Co nanoparticles as shown in fig.1b. The magnetization done on both the samples show a negligible hysteresis at 295 K and the nanospheres behave similar to superparamagnetic moments. Enormous increase in coercive field $H_C$, from 295 K to 5 K by a factor of 7 and 13 in more oxidized (fig.4a) and in less oxidized nanospheres (fig.4b) respectively is understood by the arrest of thermally assisted hopping, between different magnetic orientations, at 5 K.[4,18-19] An exchange bias field of $H_E$ = 132 Oe and 483 Oe appeared in more oxidized and less oxidized nanospheres as shown in fig.4 when the samples were field-cooled in 0.5 Tesla. Inspite of smaller antiferromagnetic shell, a substantial exchange bias for the field cooled case is seen in less oxidized nanospheres because of CoO shell. Magnetization measurements on Co nanocubes revealed a very small coercive field at 10 K as shown in fig.5b. No exchange bias effect was found when the samples were field cooled in 2 Tesla field as shown in inset fig.5b, which confirms the stable and un-oxidized nanoparticles. The blocking temperature was estimated to be 122 K and 150 K for more and less oxidized nanospheres, whereas no such behavior is noticed in other samples up to measurement temperature 350 K.

In order to quantify the effect of shape and size precisely, we have measured key magnetic parameters, like coercivity $H_C$, and remanence ratio (defined as ratio of remanent magnetization $M_r$ to saturation magnetization $M_S$), from the hystersis loop as a function of shape and size of Co nanoparticles. These are listed in Table 2. These three parameters for cube shaped nanoparticles are comparatively smaller than spherical oxide coated nanoparticles. High energy loss in alligning the magnetic moments in oxidized nanospheres compared to other nanostructures is understood as the requirement of extra energy to rearrange the interfacial spins. Relatively large remanence ratio with small hysteresis present in spherical particles is of importance for technological applications. The saturation magnetization of all the samples is small as compared to the bulk magnetization of hcp Co ($M_S \sim 162$ emu/g)[20] because of significant increase in the volume fraction of surface atoms within the whole particle and the addition of ligands at the surface of the particles.[21] The reduction of $M_S$ in case of oxidized spherical particles is also because of the presence of antiferromagnetic layer. The maximum value of $M_S$ obtained was about 78% in case of C45, and it decreased as the size of the cubes decreased as shown in fig.5b, because of surface effects. The same behavior is observed in NS and S+D nanoparticles as shown in fig.5a. The electronic and magnetic configuration of nanoparticles are dramatically influenced by the absorption of the surfactant ligands at the surface of nanoparticles that alters the surface magnetic structure and thereby impacts on the magnetic properties of the system.[22-23] Given the major differences in the magnetic properties of the samples which contain Co nanodiscs, magnetic separation is the easiest method to isolate these particles after the reaction is quenched.

\*\* Table 2

In conclusion cobalt nanoparticles of dissimilar morphologies were prepared by thermal decomposition of cobalt carbonyl. Surfactants played a major role in controlling the shape and size of the Co nanoparticles. The properties of linear amine surfactant influenced the particle shape and lead to flat shape. Magnetic characterization revealed that shape anisotropy played an important role in controlling the magnetic properties of Co nanoparticles. Co nanospheres show strong ferromagnetic behavior at low temperatures whereas nanocubes have very small coercive field. Oxidization of nanospheres gave rise to exchange bias and high remanence ratio at low temperatures.

**Acknowledgement**

DSK acknowledges UGC for financial support. DST, Government of India is acknowledged for funding the VSM at CSR, Indore and High field magnet facility at JNU, New Delhi. SBP thanks DST for funding under Young Scientist Scheme.

**Figure Caption**

Fig.1. Bright field HRTEM images of oxidized nanospheres with average diameter 14.5 nm. Inset of (a) shows more oxidized nanosphere with core diameter 8.7 nm and $Co_3O_4$ shell thickness 2.5 nm. (b) Electron diffraction micrograph rings match with *hcp* phase of Co and (c) less oxidized nanospheres with inset showing a particle having core diameter 10.6 nm and shell of CoO having thickness 1.9 nm.

Fig.2. (a) Nearly spherical nanoparticles with average size 10.5 nm synthesized with low concentration of TOPO (b) spherical and disc shaped nanoparticles formed in the presence of ODA.

Fig.3. Nanocubes of (a) average length 44.5 nm obtained for a reaction time of 3 min and (b) ~ 30 nm for a reaction time of 1 min. (c) cube and disc shaped nanoparticles formed in the presence linear amine. Inset shows microscope image taken at 20° tilt and (d) single crystalline nature of nanocubes.

Fig.4. Hysteresis loops of oxidized nanospheres measured at different temperatures. Zero-field cooled and field-cooled (0.5 T) magnetization curves at 5 K exhibiting exchange bias field $H_E$ (a) 132 Oe and (b) 483 Oe in more oxidized and less oxidized nanospheres.

Fig.5. Hysteresis loops at 10 K for (a) nearly spherical nanoparticles and a combination of spherical and disc shaped nanoparticles, (b) Nanocubes exhibiting small coercivity, and inset shows zero field cooled and 2 T field cooled plot of nanocubes and discs without exchange bias effect.

* Table 1. Summary of precursor solution, surfactant mixtures employed to yield different Co nanostructures and are named for future references:

| Figure no. | Precursor solution | Surfactants | Reaction time (min) | Shape | Average size (nm) |
|---|---|---|---|---|---|
| 1(a) | 0.40 g $Co_2(CO)_8$ in 3.6 mL DCB | 0.2 g TOPO  0.1 mL OA | 15 | Sphere | 14.5 |
| 1(b) | 0.40 g $Co_2(CO)_8$ in 3.6 mL DCB | 0.3 g TOPO  0.1 mL OA | 25 | Sphere | 14.5 |
| 2(a) | 0.54 g $Co_2(CO)_8$ in 3 mL DCB | 0.1 g TOPO  0.2 mL OA | 5 | Nearly Spherical (NS) | 10.5 |
| 2(b) | 0.54 g $Co_2(CO)_8$ in 3 mL DCB | 0.9 g ODA  0.1 mL OA | 10 | Sphere and disc (S+D) | 11.5  3.8 × 13.5 |
| 3(a) | 0.62 g $Co_2(CO)_8$ in 4 mL DCB | 0.54 mL Jeffamin D-400  0.09 mL Octanic acid  0.12 g TOPO | 3 | Cube (C45) | 44.5 |
| 3(b) | 0.62 g $Co_2(CO)_8$ in 4 mL DCB | 0.54 mL Jeffamin D-400<br><br>0.09 mL Octanic acid<br><br>0.12 g TOPO | 1 | Cube (C30) | 30 |
| 3(c) | 0.54 g $Co_2(CO)_8$ in 3 mL DCB | 0.6 g ODA<br><br>0.2 g TOPO | 3 | Cube and disc (C+D) | 19.6<br><br>4 × 19.3 |

** Table 2. Macroscopic magnetic parameters like coercivity $H_C$, saturation magnetization $M_S$ at 4 T field, remenance $M_r/M_S$ calculated from the magnetization measurements:

| Shape (size in nm) | $H_C$ (Oe) | $M_S$ (emu/g) | $M_r/M_S$ |
|---|---|---|---|
| Sphere (more oxidized) (14.5) | 460 | 84.6 | 0.35 |
| Sphere (less oxidized) (14.5) | 767 | 68.7 | 0.37 |
| Nearly spherical | 500 | 64.3 | 0.35 |
| Sphere+disc | 408 | 78.3 | 0.32 |
| Cube (44.5) | 249 | 127 | 0.17 |
| Cube (30) | 236 | 54.6 | 0.11 |
| Cube +disc | 314 | 53.5 | 0.19 |

**Fig.1**

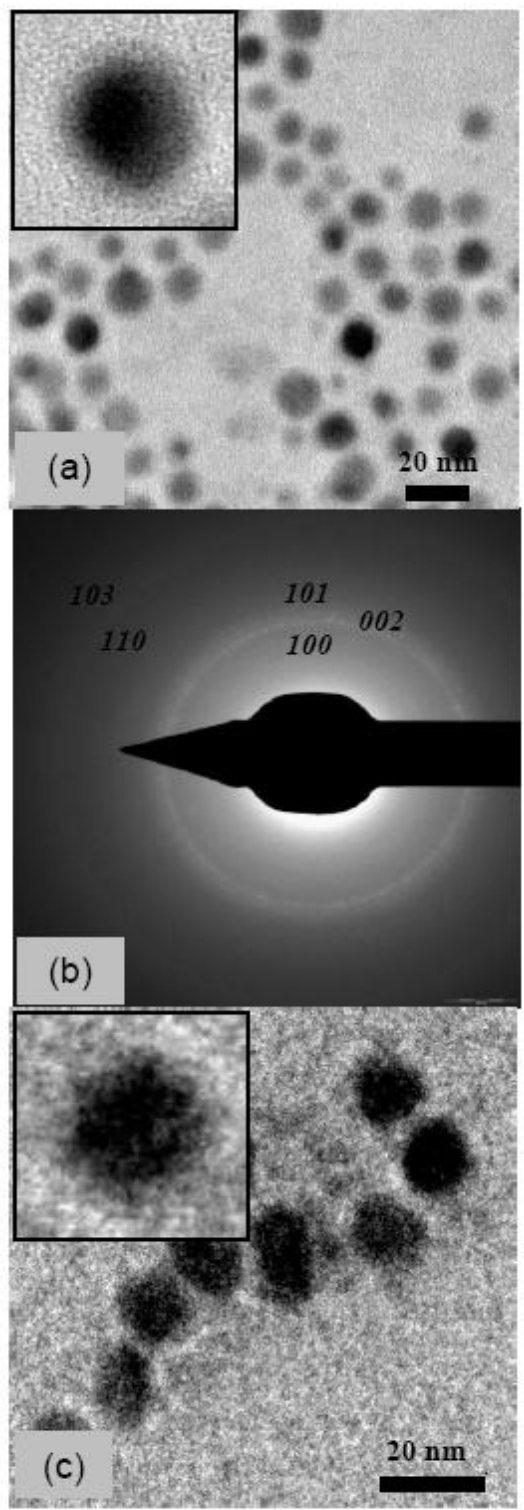

**Fig.2**

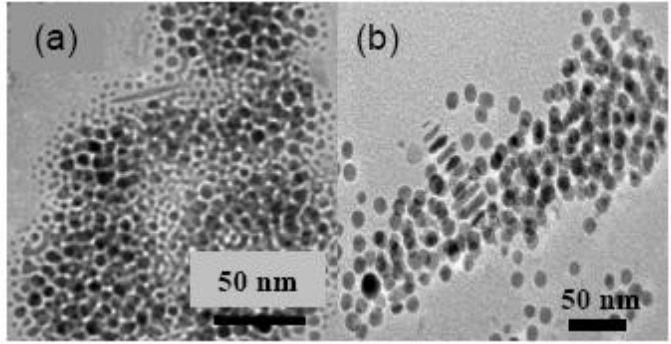

**Fig.3**

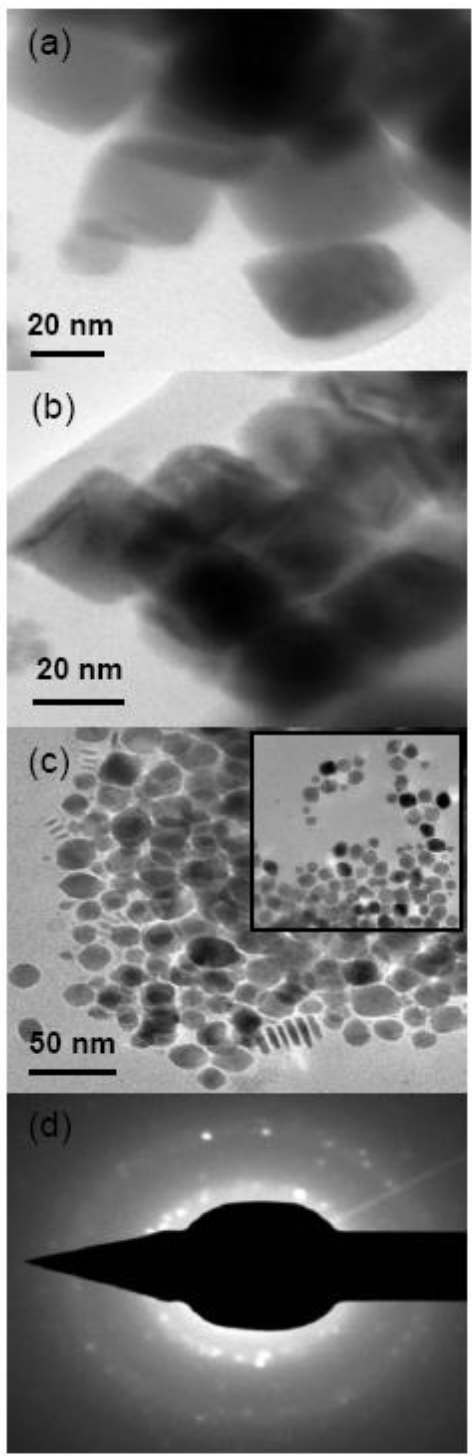

**Fig.4**

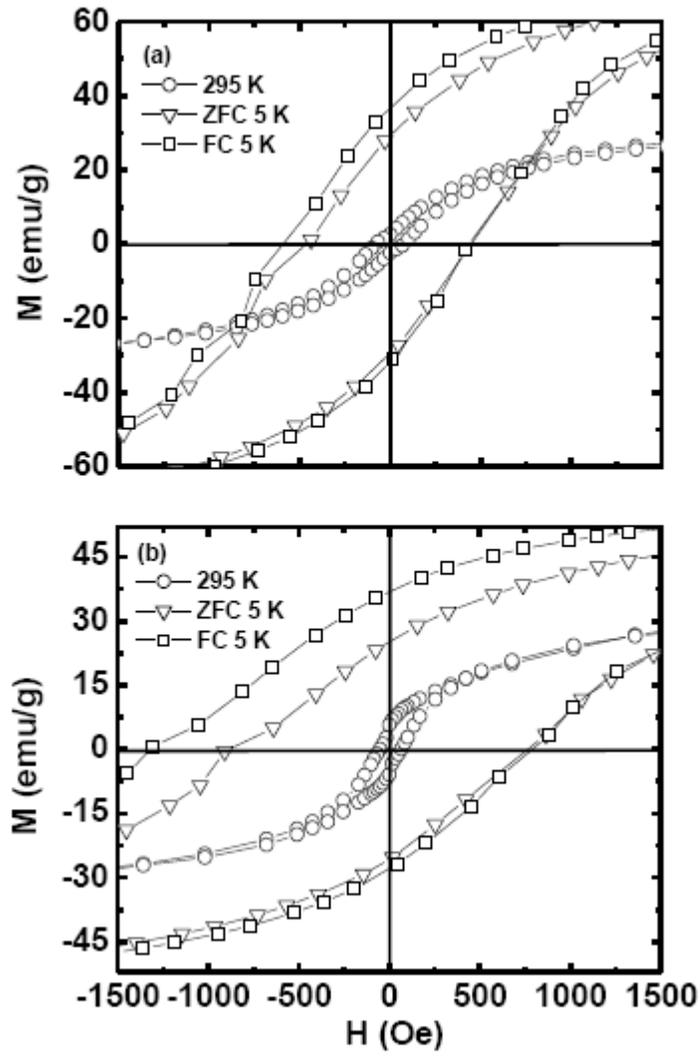

**Fig.5**

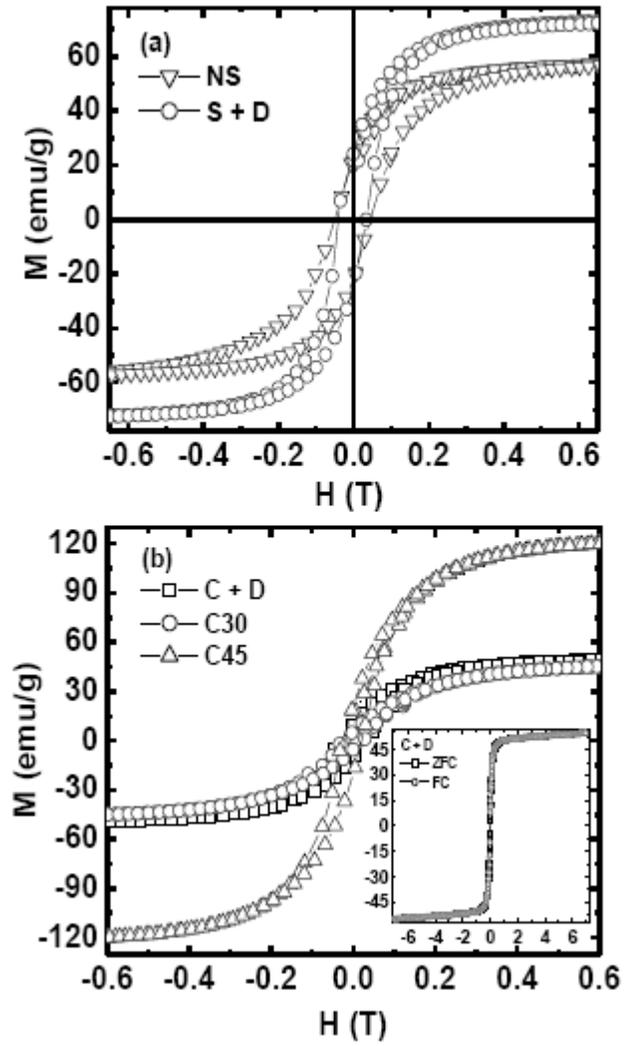